\documentclass[aps,pre,showpacs,twocolumn,superscriptaddress,unsortedaddress,floatfix]{revtex4}
\usepackage{epsfig}
\usepackage{graphicx}
\usepackage{bm}
\usepackage{amsfonts}
\usepackage{amssymb}
\usepackage{amsmath}
\usepackage{dcolumn}
\usepackage{latexsym}

\begin{document}

\title{Calculation of the elastic scattering parameters in an ultra-cold
Fermi-Bose and Bose-Bose Fr vapor}
\author{M. Kemal \"{O}zt\"{u}rk}
\email{ozturkm@gazi.edu.tr}
\author{S\"{u}leyman \"{O}z\c{c}elik}
\email{sozcelik@gazi.edu.tr}
\affiliation{Physics Department, University of Gazi, 06500 Teknikokullar, Ankara, Turkiye}
\date{Today}

\begin{abstract}
The calculations of the elastic scattering properties of ultracold-francium
isotopes are reported in the detail. A parametric model potential for the
triplet molecular states of the Fr$_{2}$ is represented, and the scattering
lengths at and the effective range $r_{e}$ are calculated using WKB and
Numerov methods for Fr-Fr in the triplet state. The convergence of these
scattering properties as the depending on a $K_{0}$ parameter and core
radius is also investigated using two method as Quantum Defect Theory and
analytical formula of symskovisky.
\end{abstract}

\maketitle

\section{\textbf{INTRODUCTION}}

\label{int} In the Bose-Einstein condensation of alkali metal atoms, about
francium atom has been known very little knowledge except trapping and
cooling until now \cite{1}. Because it has too much energy coming from
nuclear reaction and no stable long-lived isotopes. Ultracold atomic
collisions gained renew an important have came out a necessity of research
as result of the atomic trapping and cooling when energy levels, hyperfine
structure intervals and isotopes shifts for Fr isotopes is intensely studied
in detailed\cite{2,3}. Calculations of ultracold elastic scattering
properties for this atom haven't been seen in literature until now. Reason
of this is that experimental work on Fr has been lacking mostly due to no
stable long-lived isotope\cite{4,5}. Collisions of alkali metal atoms except
Fr atom at a few mill degrees Kelvin temperatures have been studied recently
in many experiments \cite{6,7,8,9,10,11}. Theoretical work about scattering
on Fr has been scant and only a few published papers exist\cite{12}.

The character of the interaction between atoms in low-temperatures is
determined by signs $(\pm)$ and magnitude of the scattering length $a$: for
bosonic and fermionic atoms with $a>0$ Bose condensate is stable, for $a<0$
it is unstable\cite{13}. The Pauli Exclusion Principle (which was determined
experimentally) states that no two fermion particles can occupy the same
state at the same time. They must have some way of being distinguished,
whether by location, spin state, or some other property. That means that if
one fermion is in a local ground or minimum energy state, the next fermion
in the area must be in a higher energy state. For bosons, however, the Pauli
Exclusion Principle is irrelevant by definition -- so all of the bosons can
be in the same state at the same time. They don't have to be distinguishable
from each other. When this happens, a Bose-Einstein Condensate is stable.
Accurate calculations of s-wave scattering length and effective range for
diatomic potentials are important due to the elastic collisions in these
temperatures dominated by s-wave or low wave scattering. The accurate
determination of scattering length depends on choosing the interaction
potentials. Collision processes at near- zero temperatures $(\sim\mu K)$ are
sensitive to the details of the interaction potentials between the colliding
systems.

In this paper, we have calculated scattering lengths $a$, effective ranges $%
r_{e}$ and cross section for collisions of Fr-Fr isotopes at ultracold
temperatures with the semiclassical method (WKB)\cite{14} and also solved
the Schr\"{o}dinger equation using Numerov method numerically\cite{15}. The
interaction of francium isotopes in triplet $(^{3}\sum_{u}^{+})$ is
described by short and long range potentials. This interaction potential $%
\mathit{V(r)}$ for Fr atom was parametered by using coefficients found in
the literature. These calculated scattering properties using this potential
are compared for two methods. We investigated the convergence of some
scattering properties obtained with help of these methods as the dependence
of the core radius and a $K^{0}$ parameter using Quantum Defect Theory\cite%
{16} and the analytic calculations of the scattering lengths obtained by
Szmytkowski\cite{17}.

\section{\textbf{PROCESSES}}

\label{processes}

The scattering length is defined from the asymptotic behavior of the
solution of the radial Schr\"{o}dinger equation at zero energy:

\begin{equation}
\left[ {\frac{d^{2}}{dr^{2}}-2\mu V(r)+k^{2}+\frac{l(l+1)}{r^{2}}}\right]
\text{y}_{l}\text{(r)=0}\mathit{\quad}\quad\quad  \label{eq1}
\end{equation}
in atomic unit. Asymptotic behavior of the wave function is

\begin{equation}
y(r)=s{\ }r+s_{0},  \label{eq2}
\end{equation}
as $r\rightarrow\infty$, where $s$ and $s_{0}$ are constants. The scattering
length is given by

\begin{equation}
a=-\frac{s}{s_{0}},  \label{eq3}
\end{equation}
here coefficients are obtained by using WKB method from semiclassical
behavior and with help of the exact solution at zero-energy. eq. ~\ref{eq3}
is transformed to the form \cite{18}:

\begin{equation}
a=\bar{a}\left\{ {1-\tan\{\varphi-\pi/8\}}\right\}  \label{eq4}
\end{equation}
in atomic unit, where $\bar{a}$ is the \textquotedblleft
mean\textquotedblright\ or \textquotedblleft typical\textquotedblright\
scattering length determined by the asymptotic behavior of the potential
through the parameter $\gamma=\sqrt{2\mu\kappa}$ ($\mu$ is reduced mass, and
$\kappa=c_{6}$ is Van der Waals constant) for atom-atom interactions:

\begin{equation}  \label{eq5}
\bar{a} = \sqrt{2\gamma} \Gamma(3 / 4) / \Gamma(1 / 4)
\end{equation}

Scattering length also depends a semiclassical phase $\varphi$ calculated at
zero energy from classical turning point $r_{0}$ where $y(r_{0})=0$, to
infinity,

\begin{equation}  \label{eq6}
\varphi= \int\limits_{r_{0} }^{\infty}{\sqrt{ - 2\mu V(r)} dr} .
\end{equation}

\begin{figure}[tbp]
\includegraphics[width=8cm]{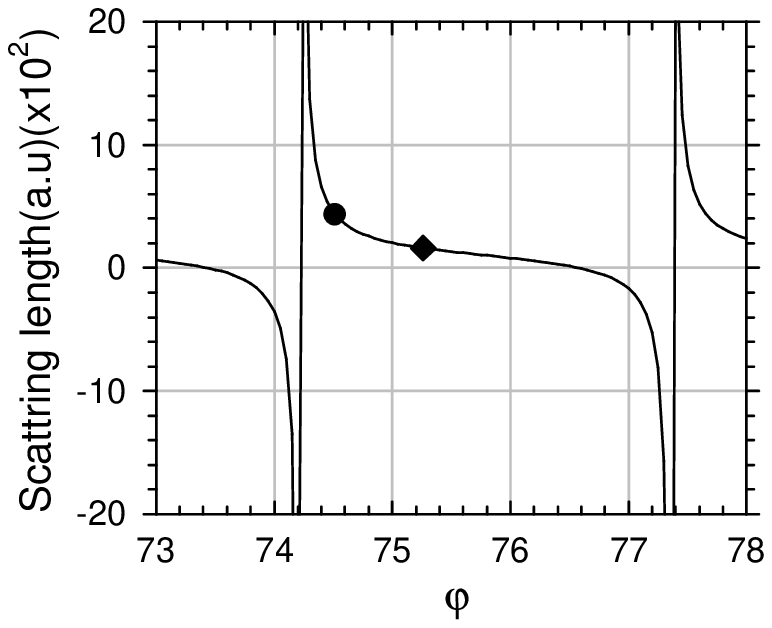}\newline
\caption{Changes of scattering length as a function of the semiclassical
phase. The solid point in the left shows triplet scattering length for $%
^{213}$Fr - $^{213}$Fr isotope and other is the result of 1{\%} changes in
phase.}
\label{fig1}
\end{figure}

It also determines the total number of vibrational levels with zero orbital
angular momentum, $N_{bd}=\{\varphi/\pi-5/8\}+1$, where {\{} {\}} is the
integer part. When the difference in brackets is just below an integer the
scattering, eq. \ref{eq4} is anomalously large negative (see figure \ref%
{fig1}), $\left\vert a\right\vert >>\bar{a}$, which correspond to presence
of a virtual level at $\varepsilon=\hbar^{2}/(2\mu a^{2})$, and when it
exceeds an integer by a margin, $a$ is very large positive, due to the
existence of a weakly bound state. Unlike $\gamma$ and $\bar{a}$, the phase
factor $\varphi$ depends strongly on the actual shape of the interatomic
potential well. When the phase is large, $\varphi/\pi>>1$ , the scattering
length is very sensitive to the slightest changes of the potential\cite{18}.
It shows clearly that in figure \ref{fig1}, the change of 1{\%} percent in
phase gives the less scattering length as 161.238a.u. instead of 434.477a.u.
for $^{213}$Fr-$^{213}$Fr isotope interaction. So, the accurate
determination of potential is very important for scattering properties.

The effective range $r_{e}$ can be written in terms of zero-energy solutions
of the partial-wave equation if $u_{0}(r)$ is the solution of the
partial-wave equation at $k=0$ for zero-potential, and normalized as

\begin{equation}
u_{0}=\frac{\sin(kr+\delta_{0})}{\sin\delta_{0}}  \label{eq7}
\end{equation}
as $k\rightarrow0$, and if $u_{0}(r)$ is the normalized solution of the
radial equation \ref{eq1} for the s partial wave at zero energy (for
non-zero potential) with the boundary condition $u_{0}(r\rightarrow\infty)=0$
as $u(r)\sim u_{0}(r)$ at $r\rightarrow\infty$. Then, effective range can be
written as \cite{19}

\begin{equation}
r_{e}=2\int\limits_{0}^{\infty}{\left\{ {u_{0}^{2}(r)-u}^{2}{(r)}\right\} dr}%
.  \label{eq8}
\end{equation}

This integral converges provided $u(r)$ approaches $u_{0}(r)$ rapidly enough
as $r\rightarrow\infty$. This requires $V(r)$ to decrease faster than $%
r^{-5} $. For $-c_{6}/r^{n}$potential type, this expression is adjusted by
Flambaum et al.\cite{13} using WKB method and the exact solution of the
radial equation at zero--energy to form

\begin{equation}
r_{e}=\frac{\sqrt{2\gamma}}{3}\left[ {\frac{\Gamma\left( {1/4}\right) }{%
\Gamma(3/4)}-2\frac{\sqrt{2\gamma}}{a}+\frac{\Gamma(3/4)}{\Gamma(1/4)}\frac{%
4\gamma}{a^{2}}}\right] .  \label{eq9}
\end{equation}
Then, the mean effective range may be rewritten by replacing $a\sim\bar{a}$
into eq. \ref{eq9} as

\begin{equation}  \label{eq10}
\overline{r_{e} } = \frac{\sqrt\gamma}{3}\left[ {\frac{\sqrt2 \Gamma\left( {%
1 / 4} \right) }{\Gamma(3 / 4)} - \frac{4\sqrt\gamma }{\overline a } + \frac{%
\Gamma(3 / 4)}{\Gamma(1 / 4)}\frac{4\sqrt2 \gamma }{\overline a ^{2}}} %
\right] .
\end{equation}

The low-energy scattering is dominated by the contribution $l=0$. At values
of $k$ close to zero, the $l=0$ phase shifts $\delta _{0}$ can be
represented by a power series expansion in $k$ \cite{20}

\begin{equation}  \label{eq11}
k\cot\delta_{0} = - \frac{1}{a} + \frac{1}{2}r_{e} k^{2} + o\left( {k^{3}}
\right) .
\end{equation}

\section{\textbf{RESULTS AND DISCUSSIONS}}

\label{result}

\subsection{\textbf{Potential}}

\label{pot} The interaction between Fr atoms like the other alkali metal
atoms has two features: First, the potential at large distances behave as an
inverse power of the interatomic distance, $\upsilon =-\kappa /r^{n}$ with $%
n=6$ for spherically symmetric two atoms in their ground state. The
asymptotic parameter $\kappa \equiv c_{6\text{ }}$is known quite well for
most atomic pairs of interest. Second, for alkali metal atoms other than
hydrogen and helium the potential curve is generally quite deep and also the
electron-exchange part of the atomic interaction is repulsive at smaller
distance as for the singlet and triplet terms of alkali- metal atoms.
\textquotedblleft Deep\textquotedblright\ here means that the wave functions
of the atomic pair oscillate many times within the potential well, even at
very low collision energies. The interatomic potential supports a large
number of vibration levels. This latter feature enables one to use the
semiclassical approximation to describe the motion of Fr atoms within
potential well \cite{21}.

Alkali-metal atoms have uncompensated electron spin $s=1/2$ in the ground
state. Therefore, the interaction between two alkali atoms in the ground
state results in the formation of a diatomic molecule, which is described by
two terms corresponding to the total spin $s_{ab}=s_{a}+s_{b}$ of the system
equals to 1 or 0. The state with $s_{ab}=0$ and 1 corresponds to the singlet
state and triplet state, respectively. The probability of the spin exchange
is determined by both the dependence of potential curves on the internuclear
distance $r$ and the splitting between singlet and triplet terms. The
exchange interaction is analytically calculated using the surface integral
method by Smirnov and Chibisov\cite{22}, which yields

\begin{equation}
v_{exc.}(r)=\frac{1}{2}(v_{u}-v_{g})=J(A,\alpha,\beta)r^{\frac{2}{\alpha }+%
\frac{2}{\beta}-\frac{1}{\alpha+\beta}-1}e^{-(\alpha+\beta)r}.  \label{eq12}
\end{equation}

Where $\alpha=\beta$ is the ionization energy of francium atom, and $J$ are
constant. In this equation, $J$ is calculated from the surface integral
given in the Ref.\cite{22}. Also we have obtained the value of normalization
constant, A by joining in the range of distances from the nucleus the wave
function $y(r),$ obtained by the Hartree-Fock method, and the asymptotic
value of the wave function of Eq. (\ref{eq2}) for an electron far from the
nucleus but in the Coulomb field of the atomic core. For Fr-Fr, these values
of constants are given in Table \ref{tab1}.
\begin{table}[ptbh]
\caption{Potential coefficients for Francium-Francium atom in triplet state.}
\label{tab1}%
\begin{tabular}{ccccc}
\hline\hline
A & J & $\alpha$ & C$_{3}$ & C$_{6}$ \\ \hline
0.500 & 4.480x10$^{-3}$ & 0.5471 & 4.437 & 5256 \\ \hline\hline
\end{tabular}%
\end{table}

The long-range part of the interaction potential is given by

\begin{equation}
v_{long-range}=-\mathop\Sigma\limits_{m}^{even}\frac{C_{m}}{r^{m}}f_{c}(r),
\label{eq13}
\end{equation}

which is the sum of Van der Waals terms. The $C_{3}$ and $C_{6}$ Van Der
Waals values are given in table \ref{tab1} had been calculated by Derivianko
in literature\cite{23}. The potential of Fr+Fr atom interaction consists of
above expressed two terms, the sum of Eqs. \ref{eq12} and \ref{eq13}, is
given as

\begin{equation}
V^{2s_{AB}+1}(r)=V(r)_{long-range}+(-1)^{s_{ab}}V(r)_{exc.}  \label{eq14}
\end{equation}
Here, for triplet state, $S_{AB}$ is taken as 1 value. This potential can be
usefull for other studies.

\subsection{Scattering length and effective range of Fr}

\label{scatt}

The scattering lengths of francium isotopes calculated from Eq.\ref{eq4}
using WKB method for $V(r)$ are given in the Table \ref{tab2}. In order to
calculate numerically scattering length, we obtained s-wave phase shift $%
\delta_{0}$ using the potentials in eq.\ref{eq14}, then we do it by solving
radial equation using Numerov algorithm at $e=k^{2}/2\mu$ in atomic unit and
finding $\delta_{0}$ from the asymptotic behavior of the wave function $%
y(r)=\sin (kr+\delta_{0})$. The phases at small momenta $k$ are used to
extract the scattering length numerically from $a=%
\mathop{\lim (\tan
\delta _{0}/k)}\limits_{k\rightarrow0}$. The scattering lengths obtained via
this method are given in the numeric columns of Table \ref{tab2}.

\begin{table}[ptbh]
\caption{Scattering lengths of the samespecies francium isotope interactions
in the triplet state. x,y gives isotopes cases.}
\label{tab2}%
\begin{tabular}{ccc}
\hline\hline
\multicolumn{3}{c}{Triplet scattering lengths $a_{t}(a.u.)$} \\ \hline
Isotopes($^{x}$Fr-$^{y}$Fr) & WKB & Numerov \\ \hline
$^{208-208}$ & -52.3151 & -52.303 \\
$^{209-209}$ & -134.209 & -134.229 \\
$^{210-210}$ & -332.242 & -332.672 \\
$^{211-211}$ & -1779.55 & -1779.490 \\
$^{212-212}$ & 932.576 & 932.598 \\
$^{213-213}$ & 434.477 & 434.481 \\
$^{214-214}$ & 301.464 & 301.461 \\
$^{215-215}$ & 237.152 & 151.911 \\
$^{216-216}$ & 197.235 & 197.226 \\
$^{223-223}$ & 52.0355 & 52.035 \\ \hline\hline
\end{tabular}%
\end{table}

Some parameters are important in the determination of $a$ \ref{eq15} for WKB
method. We also obtain these parameters, the mean scattering length $\bar{a}$
and the zero-energy semiclassical phases $\varphi$, given in Table \ref{tab3}%
, using values of asymptotic parameter $\gamma$. Here, the mean scattering
length with asymptotic behavior of the potential in Eq.(\ref{eq14}) has been
calculated from equation (\ref{eq5})

\begin{table}[ptbh]
\caption{Mean scattering length $\bar{a}$, semiclassical phase shifts $%
\protect\varphi$ and numbers of the vibration bound states of samespecies
isotope interactions in the triplet states.}
\label{tab3}%
\begin{tabular}{cccc}
\hline\hline
($^{x}$Fr-$^{y}$Fr) & $\bar{a}_{WKB}$ & $\varphi$ & $N_{b}$ \\ \hline
$^{208-208}$ & 101.173 & 73.6373 & 22.8145 \\
$^{209-209}$ & 101.294 & 73.8139 & 22.8707 \\
$^{210-210}$ & 101.415 & 73.9904 & 22.9269 \\
$^{211-211}$ & 101.536 & 74.1662 & 22.9828 \\
$^{212-212}$ & 101.656 & 74.3419 & 23.0387 \\
$^{213-213}$ & 101.775 & 74.5170 & 23.0945 \\
$^{214-214}$ & 101.895 & 74.6922 & 23.1503 \\
$^{215-215}$ & 102.014 & 74.8667 & 23.2058 \\
$^{216-216}$ & 102.133 & 75.0412 & 23.2613 \\
$^{223-223}$ & 102.952 & 76.2502 & 23.6462 \\ \hline\hline
\end{tabular}%
\end{table}

As shown in the Table \ref{tab2}, the scattering lengths calculated for the
collisions of Francium pairs is in excellently agreement for two method.

Even our results shows that the stability of large condensates requires
repulsive interactions (positive $a)$, whereas for attractive interactions
(negative $a)$ it is unstable, only a finite number of atoms can be found in
condensate state in a trap. Due to changes in the V(r), scattering length is
found as negative value, which lead to a condensate triplet state where the
number of atoms is limited to a small critical value determined by the
magnitude of $a$ . In contrast, we have observed the positive scattering
lengths that produce stable condensates The calculations given in Table \ref%
{tab2} shows that all atoms are found in the same energy levels If
interaction atom species are boson, or they display a adopt behavior due to
Pauli exclusion principle If interaction atom species are fermion. Also,
from scattering lengths given as a function of the semiclassical phase shift
in the eq. (\ref{eq4}), we say that the minor changes in phase shifts have
strongly affected the scattering amplitude. It is also remarked that for $%
^{212}$Fr-$^{212}$Fr, the analysis in the direct determination of scattering
length is weakly dependent on the number $N_{b}$ of bound states supported
by ab-initio potential. In contrast, mass scalling from one isotope to the
others depends more strongly on $N_{b}$.

As shown in table \ref{tab2} scattering lengths are unstable against
collapse from isotope $^{208}$Fr-$^{208}$Fr to $^{211}$Fr-$^{211}$Fr. Here
the both boson-boson and fermion-fermion interactions are unstable. Same
case is valid for the other stable isotope interactions.

Also, the effective ranges calculated using Numerov and WKB methods in the
function of phase for the $V(r)$ potentials are presented in Table \ref{tab4}%
. The close agreement between the calculations of $r_{e}$ from Eqs.(\ref{eq8}%
) and (\ref{eq11}) confirms the accuracy of the numerical integration of the
partial-wave equation. The size of the scattering lengths and effective
ranges is closely related to the position of the last vibration bound states
of the energy curves, as can be anticipation by inspection of Eq.(\ref{eq3})
and number of vibrational levels $n_{s}(\varphi/\pi)$ which, consistent with
Levinson's theorem, show that as the binding energy of the highest level
tends to zero, the scattering length tends to $\pm$ infinity\cite{24}.

\begin{table}[ptbh]
\caption{Effective ranges of the samespecies francium isotope interactions
in the triplet state. x,y gives isotopes cases.}
\label{tab4}%
\begin{tabular}{lll}
\hline\hline
\multicolumn{3}{c}{Effective ranges $r_{t}$} \\ \hline
($^{x}$Fr-$^{y}$Fr) & $r_{WKB}$ & $r_{Numerov}$ \\ \hline
$^{208-208}$ & 3645.27 & 3635.894 \\
$^{209-209}$ & 1078.46 & 1078.008 \\
$^{210-210}$ & 531.721 & 531.615 \\
$^{211-211}$ & 332.01 & 332.150 \\
$^{212-212}$ & 239.005 & 239.243 \\
$^{213-213}$ & 190.433 & 190.127 \\
$^{214-214}$ & 164.266 & 164.168 \\
$^{215-215}$ & 151.738 & 151.738 \\
$^{216-216}$ & 149.197 & 149.197 \\
$^{223-223}$ & 1463.56 & 1463.560 \\ \hline\hline
\end{tabular}%
\end{table}

\subsection{Convergenge of scattering length and effective range}

\label{con} Analytical calculations of scattering lengths are important in
investigation of convergence of the scattering length and the effective
range. In many approaches used to solve collision problems in atomic
physics, the tree dimensional configuration space is divided into two
regions separated by a spherical shell (a core boundary) of radius $\rho$.
In the inner region $(r<\rho)$ the short-range interaction between two
colliding particles is very complicated and a scattering equation must be
solved independently for each combination of particles. In contrast, it $%
\rho $ is chosen sufficiently large the scattering problems in the outer
region $(r>\rho)$ may be reduced to potential scattering with the long-range
potential accurately approximated by simple analytical expression. A
numerical solution in this region is usually easily approachable\cite{25}.
Two of these methods are analytical calculations derived by Radoslaw
Szmytkowski for inverse problems\cite{25} and Quantum Defect Theory
described by Gao\cite{16}.

Radoslaw Szmytkowski solved the radial schr\"{o}dinger equation at zero
energy for inverse type of long-range interaction analytically with
solutions expressed in terms of the Bessel, Whittaker and Legendre
functions, respectively. They derive exact analytical converge formulae for
the scattering lengths. The expressions depend on the short-range scattering
length, core radius and parameters characterizing the long-range part of the
interaction. In Table \ref{tab5}, We obtained the convergence of the ultra
cold scattering length of Francium atom using the formulae of Szmytkowski.
These formulae are given in that paper as eqs. (75), (73). As shown in the
table, this method confirms convergence of the scattering length for $^{212}$%
Fr-$^{212}$Fr. Here, only, the convergence of the scattering length
calculated from WKB method is given, and the same behavior can also be seen
in the numerical calculation.

\begin{table}[ptbh]
\caption{Convergence of scattering length calculated by WKB with the
formulae of Szmytkowski for $^{212}$Fr-$^{212}$Fr.}
\label{tab5}%
\begin{tabular}{ccc}
\hline\hline
$\rho$ & $a_{0}$ & $a_{02}$ \\ \hline
10 & -73.2943 & -6.19E+09 \\
100 & 53.1953 & -46796 \\
500 & 929.775 & 929.775 \\
1000 & 932.426 & 932.426 \\
5000 & 932.569 & 932.569 \\
10000 & 932.575 & 932.575 \\
50000 & 932.576 & 932.576 \\
Infinity & 932.576 & 932.576 \\ \hline\hline
\end{tabular}%
\end{table}

Quantum defect theory of atomic collisions is presented by Gao. Based on the
exact solutions of the Schr\"{o}dinger equation for an attractive $1/r^{6}$
potential, the theory provides a systematic interpretation of molecular
bound states and atom-atom scattering properties and establishes the
relationship between them. He finds a definition for the scattering length
and the effective range and s wave at zero energy as

\begin{equation}
\left. {%
\begin{array}{l}
a_{l=0}=\frac{2\pi }{[\Gamma (1/4)]^{2}}\frac{K^{0}-1}{K^{0}}\beta _{6} \\
\\
r_{l=0}=\frac{[\Gamma (1/4)]^{2}}{3\pi }\frac{\left[ {K^{0}}\right] ^{2}+1}{%
\left[ {K^{0}}\right] ^{2}}\beta _{6} \\
\end{array}%
}\right\}  \label{eq15}
\end{equation}
where $\beta _{6}=\left( {2\mu \,C_{6}}\right) ^{4}$ and $K^{0}$is the
analytic function of energy.

\begin{figure}[ptb]
\includegraphics[width=8cm]{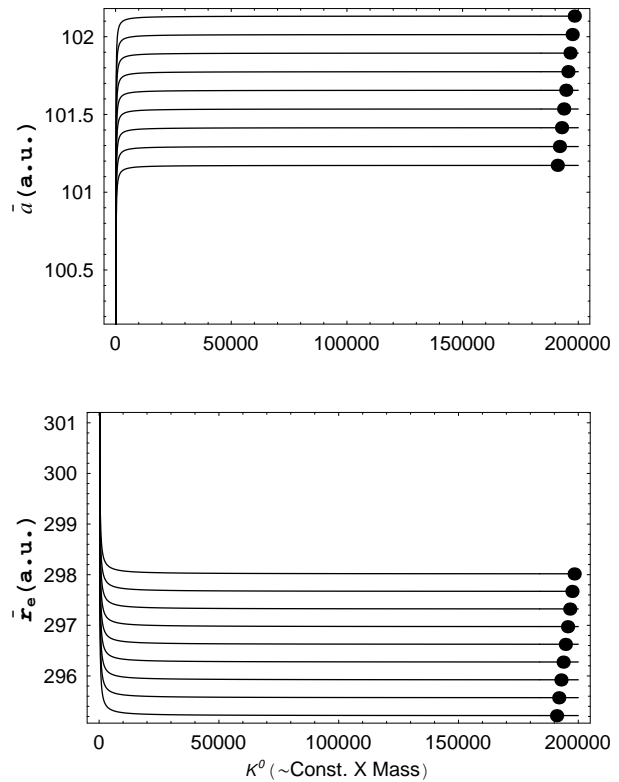}\newline
\caption{Convergence of the scattering length and the effective range
calculated with QDT as a function of $K^{0}$ parameter when $K^{0}$ goes
infinity.}
\label{fig2}
\end{figure}

Figure \ref{fig2} shows the scattering length and the effective range curve
plotted using the eqs. (\ref{eq15}) and ones(dotted points) obtained by the
mean WKB formulas.

Our graphical representations show that as $K^{0}$ goes infinity, the QDT
results converges excellently to values obtained by WKB method in Fig. \ref%
{fig2}. Even, it may be said that the different calculation of these
properties gives similar results for any ultracold atom-atom collision in
these two method.

\subsection{The zero energy cross sections}

\label{zero}

It is also known in more tex book that the low energy scattering is always
dominated by the $l=0$ partial wave the corresponding phase shift being
given by

\begin{equation}
\delta_{0}=ka  \label{eq16}
\end{equation}

The triplet elastic cross sections can be defined by\cite{24}

\begin{equation}
\sigma _{el.}^{t}=\frac{4\pi }{k^{2}}\sum\limits_{l=0}^{\infty }{(2l+1)\sin
^{2}\delta _{l}^{t}},  \label{eq17}
\end{equation}%
where $l=0$, $\sin \delta _{0}\cong \delta _{0}$ in limit of zero-energy and
Eq. \ref{eq17} is transformed using eq. ~\ref{eq16} to $\sigma
_{el.}^{t}=4\pi \,a^{2}$. This result is valid for fermion-fermion
interaction, but it is multiplied with 2 value due to symmetric wave
function If systems are bosonic one. Table \ref{tab6} shows zero
energy-cross sections are adopted for two method. It concludes that zero
energy cross sections is expected to be smaller than that of other alkali
metal atom, because of it is a heavier atom

\begin{table}[ptbh]
\caption{Triplet cross section of different francium isotopes calculated for
two method. The numbers in brackets denote multiplicative powers.}
\label{tab6}%
\begin{tabular}{ccc}
\hline\hline
\multicolumn{3}{c}{Triplet cross section($m^{2})$} \\ \hline
($^{x}$Fr-$^{y}$Fr) & WKB & Numerov \\ \hline
$^{208-208}$ & 1.92617(-16) & 1.92528(-16) \\
$^{209-209}$ & 6.33833(-16) & 6.34022(-16) \\
$^{210-210}$ & 7.76874(-15) & 7.78886(-15) \\
$^{211-211}$ & 1.11438(-13) & 1.11430(-13) \\
$^{212-212}$ & 6.12083(-14) & 6.12111(-14) \\
$^{213-213}$ & 6.64271(-15) & 6.64283(-15) \\
$^{214-214}$ & 6.39606(-15) & 6.39593(-15) \\
$^{215-215}$ & 1.97909(-15) & 8.12064(-16) \\
$^{216-216}$ & 2.73785(-15) & 2.73760(-15) \\
$^{223-223}$ & 9.52820(-17) & 9.52802(-17) \\ \hline\hline
\end{tabular}%
\end{table}

\section{\textbf{CONCLUSIONS}}

The elastic scattering properties for the collision of the francium atoms in
a limited range of moment $(k<\bar{a}^{-1}\approx 0.01a.u.$ for
francium-francium isotopes$)$ at low temperatures are sensitive to the
details of interaction potentials. Scattering properties such as scattering
length, effective range and cross section have been computed using
semiclassical and a numeric methods for the $V(r)$ potentials as dependence
of cutoff radius adjusted by comparing with the experimental potential for
ultra-cold francium-francium isotopes collision. We investigated the
convergence of these scattering properties as the depending of core radius
and $K^{0}$ parameter using Quantum Defect Theory and analytical
calculations derived by Radoslaw Szmytkowski for inverse problems. Cross
section was obtained as $\sim $ $10^{-15}$m$^{2}$ at low energy. The phase
shifts, an intermission parameter for scattering length and effective range,
has linear manner at small momenta $k<\bar{a}^{-1}$.

\end{document}